\documentstyle[12pt]{article}
\newcommand{\be}{\begin{equation}}
\newcommand{\ee}{\end{equation}}
 \newcommand{\ovl}{\overline}
\newcommand{\bea}{\begin{eqnarray}}
\newcommand{\eea}{\end{eqnarray}}
\newcommand{\ba}{\begin{array}}
\newcommand{\ea}{\end{array}}
\newcommand{\beqa}{\begin{eqnarray}}
\newcommand{\eeqa}{\end{eqnarray}}
\newcommand{\lsim}{{\;\raise0.3ex\hbox{$<$\kern-0.75em\raise-1.1ex\hbox{$\sim$}}\;}}
\newcommand{\gsim}{{\;\raise0.3ex\hbox{$>$\kern-0.75em\raise-1.1ex\hbox{$\sim$}}\;}}

\newcommand{\NP}[1]{Nucl. Phys.\ { #1}\ }
\newcommand{\PL}[1]{Phys. Lett.\ { #1}\ }

\newcommand{\PRL}[1]{Phys. Rev. Lett.\ { #1}\ }

\newcommand{\B}{\beta}
\newcommand{\D}{\delta}
\newcommand{\DE}{\Delta}

\newcommand{\ssu}{$SU(2)_L\times SU(2)_R\times U(1)_{B-L}\,$}
\newcommand{\sul}{$SU(2)_L$}

\newcommand{\sur}{$SU(2)_R$}
\newcommand{\matr}{\left( \begin{array}}
\newcommand{\ematr}{\end{array} \right)}

\newcommand{\dis}{\displaystyle}

\newcommand{\dfrac}[2]{{\displaystyle \frac{#1}{#2}}}
\newcommand{\juuri}{\dfrac{1}{\sqrt{2}}}

\newcommand{\lr}{{left-right symmetric model}}

\begin{document}

\mbox{}\vspace*{-1cm}\hspace*{9cm}\makebox[7cm][r]{  HU-TFT-93- 51}
\medskip

\Large

\begin{center}
{\bf Signatures of left-right symmetry at high energies\footnote{Invited talk
in {\it The 2nd Tallinn Symposium on Neutrino Physics}, October 5-8, 1993}}

\bigskip
\normalsize
J. Maalampi\footnote{e-mail: maalampi@phcu.helsinki.fi}\\{\it Department of
Theoretical Physics, University of Helsinki}\\
{\it Helsinki, Finland}\\[15pt]

\bigskip

\normalsize

{\bf\normalsize \bf Abstract}

\end{center}

\normalsize

We discuss various experimental tests of the left-right symmetric
\ssu model  possible to
perform in the next generation linear colliders. We consider
processes which provide sensitive
probes of  the basic ingredients of the model:
right-handed gauge bosons, right-handed Majorana
neutrinos, lepton number violating interactions and triplet Higgs scalars. A
supersymmetric version of the \ssu model is also studied and some distinctive
experimental tests for it is proposed. One of the main messages of
this talk is to emphasize the usefulness of the collision modes $e^-e^-$,
$e^-\gamma$ and $\gamma\gamma$ for the tests of the \lr.

\bigskip \normalsize

\section{Introduction}
The left-right symmetric model
based on the gauge group \ssu [1]
is the simplest extension of the standard
 electroweak theory involving extra charged gauge bosons. Apart from its
original motivation
of  providing a dynamical explanation for the parity violation observed in
low-energy weak
interactions,  this model differs
from the Standard  Model in another important respect:
it can explain the observed
lightness of neutrinos in a natural way. Neutrino masses are created through
the
see-saw mechanism \cite{seesaw}, according to which there are in each family a
light neutrino, much lighter than the charged fermions of the family,
and a heavy neutrino.
The anomalies measured in the
solar \cite{sun} and atmospheric \cite{atmos} neutrino fluxes seem
to indicate that neutrinos indeed should have a small but non-vanishing mass.
Furthermore, the recent observations of the COBE satellite \cite{cobe} may
indicate that there exists a
hot neutrino component in the dark matter of the Universe. The see-saw
mechanism
can account for all these phenomena, while in the Standard Model neutrinos are
massless.

Several authors \cite{Gunion}, \cite{phen} have investigated    indirect
implications of the left-right symmetry on the various low-energy phenomena
to set constraints on the parameters of the model, such as the mixings between
the new gauge bosons $W_R^{\pm}$ and  $Z_R$  associated with \sur\
and  their \sul\ counterparts $W_L^{\pm}$ and  $Z_L$.
 In the case the
gauge coupling constants $g_L$ and $g_R$ of  \sul\  and
\sur, as well as the CKM-matrix and its equivalent in $V+A$ charged current
interactions, are kept unrelated, one obtains from the charged current data
the bounds \cite{phen} $g_LM_{W_2}/g_R\gsim 300$ GeV and  $g_L\zeta/g_R\lsim
0.013$, where
$\zeta$ is the $W_L-W_R$ mixing angle. From neutral current data one can derive
the lower
bound $M_{Z_2}\gsim 400$ GeV for the mass of the new $Z$-boson and the upper
bound of
$0.008$ for  the $Z_1,Z_2$ mixing  angle. CDF experiment at Tevatron one has
recently
obtained the mass limits $M_{W_2} > 520$ GeV and $M_{Z_2} > 310$ GeV
\cite{tevatron}.

In this talk I will consider various direct tests of the left-right symmetric
model,
which  one could perform in the high energy linear colliders. The collision
energies  in
these acceleratos (CLIC, NLC, TESLA, JLC) are planned to be in
the range $\sqrt s = 0.5- 2$ TeV \cite{Wiik}.  If the masses of the new gauge
bosons are close to their present lower limits, it would  be possible to
produce  them and
directly investigate their properties.  Similarly it would be possible to
search for the
heavy Majorana neutrinos, or right-handed neutrinos, predicted by the model.
The masses of
these neutrinos are, if one believes in the see-saw mechanism, in the most
simple case of
the order of the masses of the new weak bosons. One would also be able to probe
the
symmetry breaking sector of the theory by looking for the new type of Higgs
bosons, in
particular doubly charged triplet scalars which are assumed to set the large
mass scale of
the model.

The organization of this talk will be as follows. We first give a short account
of the
basic structure of the \ssu  model. In Section 3 the pair production of weak
bosons in
$e^+e^-$ collisions will be considered. It is pointed out that the cross
section of
these  processes may be quite sensitive to the mass of heavy neutrinos. In
Section 4
we will discuss processes where the lepton number violating couplings
associated with
the triplet Higgses and the Majorana neutrinos play role. In Section 5 we will
introduce a
supersymmetric version of  the \lr\  and investigate its tests in $e^+e^-$,
$e^-e^-$,
$e^-\gamma$ and $\gamma\gamma$ collisions. Some conclusions are made in Section
6.

\section{Structure of the \lr}

The left-right symmetric models are characterized by the gauge group \ssu. The
left- and
right-handed fermions are set into doublet representations of \sul\ and \sur,
respectively.
In the following we will deal with leptons only,  which are assigned according
to
\begin{equation} \Psi_L
= \matr{c} \nu_e \\ e^-\ematr_L
= ({ 2},{ 1},-1),\hspace{10pt}
\Psi_R
= \matr{c} \nu_e \\ e^-\ematr_R
= ({ 1},{ 2},-1),
\end{equation}
and similarly for other families. The $U(1)$ quantum number is normalized in
such a way
that the electric charge $Q$ is given by $Q=T_{3L}+T_{3R}+(B-L)/2$, where
$T_{3L}=T_{3R}
=\sigma_3/2$ are the doublet representations of the neutral generators of the
$SU(2)$
subgroups.

In order to generate masses for fermions one requires at least one Higgs
bidoublet of the form
\begin{equation}
\begin{array}{c}
{\dis\Phi =\matr{cc}\phi_1^0&\phi_1^+\\\phi_2^-&\phi_2^0
\ematr = ({ 2},{ 2},0),}
\end{array}
\end{equation}
whose vacuum expectation value (VEV) is given by
\begin{equation}
\begin{array}{c}
{\dis<\Phi> =\juuri\matr{cc}\kappa_1&0\\0&\kappa_2\ematr.}
\end{array}
\end{equation}
In order to break the symmetry to the electromagnetic group $U(1)_{\rm em}$
additional
higgs multiplets with $B-L \neq 0$ are needed. To introduce at the same time
Majorana mass terms for the neutrinos we add to theory the triplet Higgses
\begin{equation}
\begin{array}{c}
{\dis\Delta_L
=\matr{cc}\Delta_L^+&\sqrt{2}\Delta_L^{++}\\
\sqrt{2}\Delta_L^0&-\Delta_L^+
\ematr = ({ 3},{ 1},2)},\\[10pt]
{\dis\Delta_R=\matr{cc}\Delta_R^+&\sqrt{2}\Delta_R^{++}\\
\sqrt{2}\Delta_R^0&-\Delta_R^+\ematr = ({ 1},{ 3},2)}
\end{array}
\end{equation}
with the VEV's given by
\begin{equation}
\begin{array}{c}
{\dis<\Delta_{L,R}>
=\frac1{\sqrt{2}}\matr{cc}0&0\\v_{L,R}&0
\ematr.}
\end{array}
\end{equation}

The left-handed triplet scalar $\Delta_L$ does not play any role in the
dynamics of the
model. Its vacuum expectation value $v_L$ is tightly constrained by the
measurements of
the mass ratio of the ordinary weak bosons  which implies $v_L^2\ll
\kappa_1^2+\kappa_2^2$. On the other hand, $v_R^2\gg
\kappa_1^2+\kappa_2^2$ in order to satisfy the lower mass limits of the new
weak bosons.

The Yukawa couplings between the leptons and the scalars are the following:
\be
{\cal{L}}_{Yu} = f \overline{\Psi}_R \Phi \Psi_L
+g\overline{\Psi}_R \tilde{\Phi} \Psi_L
+ih_L \Psi_L^TC\sigma_2\Delta_L\Psi_L
+ih_R \Psi_R^TC\sigma_2\Delta_R\Psi_R + h.c.,
\ee
where $\Delta_{L,R}=\Delta_{L,R}^i\sigma_i$ and
$\tilde{\Phi}=\sigma^2\Phi^*\sigma^2$. As one can see from this the triplet
Higgses
$\Delta_L$ and $\Delta_R$ carry  lepton number --2, and on the other hand the
$B-L$
symmetry forbids their coupling to quarks.

There are all together seven vector bosons:
$W_{L,R}^\pm = \frac{1}{\sqrt{2}}(V^1_{L,R} \pm iV^2_{L,R})$, $V_{L,R}^3$,
and $B$. We define the physical states of the bosons by the equations
\be
\matr{c}
W_L^\pm \\[5pt] W_R^\pm
\ematr
=\matr{cc}
\cos\zeta&-\sin\zeta\\[5pt]\sin\zeta&\cos\zeta\ematr
\matr{c}W_1^\pm\\[5pt]W_2^\pm\ematr,
\ee
\begin{equation}
\matr{c}V_L^3\\V_R^3\\B\ematr
=(R_{ij})\matr{c}Z_1\\Z_2\\Z_3=\gamma\ematr\hspace{20pt}(i=L,R,B).
\end{equation}

The mixing matrix $R$ can be parametrized in terms of three rotation angles.
One can determine \cite{Maalampi} their values for example by using the
experimental results
for  the electron vector and axial vector neutral current couplings
and the low-energy constraint for the $W_L-W_R$ mixing angle $\zeta\lsim
0.005$\cite{altarelli}.

{}From the left-handed and right-handed neutrino states one  can form three
types
of Lorentz-invariant mass terms: Dirac term $\ovl\nu_L\nu_R$ and Majorana terms
$\ovl\nu^c_L \nu_R$ and  $\ovl\nu^c_R\nu_L$. All these terms are realized in
the
\lr\ with the Yukawa coupling (6) and the VEVS given in eqs. (3) and (5). By
assuming $v_L\simeq 0$ one ends up with the famous see-saw mass matrix
\be
   M = \left(   \begin{array}{cc} 0 & m_D \\ m_D & m_R  \end{array}  \right).
\ee
 The eigenstates of this matrix are  two
Majorana neutrinos $\nu_1$ and $\nu_2$  with approximate masses
  $ m_1 \simeq {m_D^2}/{m_R}$ and $m_2 \simeq
m_R$.
 The left-handed and the
right-handed neutrinos are related to the mass eigenstates according to
\begin{equation}
\begin{array}{c} \nu_L=\left(\nu_{1L}\cos\eta-\nu_{2L}\sin\eta\right),\\[10pt]
\nu_R=\left(\nu_{1R}\sin\eta+\nu_{2R}\cos\eta\right).\\ \end{array}
\end{equation}
where the mixing angle $\eta$ is given by
\be
   \tan 2\eta = \frac{2m_D}{m_R}.
\ee

The masses of $W_2$ and $\nu_2$ are related as
\be
m_2\simeq\frac{h_R}{g_R}M_{W_2}.
\ee
Most naturally the heavy neutrino and the heavy weak boson would
have roughly the same mass, but depending on the actual value of the Yukawa
coupling constant $h_R$ the neutrino may be much lighter or somewhat heavier
than $W_2$.

\section{Production of heavy weak bosons}

The pair production reactions
\bea
e^+e^-\to& W_2^+W_2^-\nonumber\\
\to& W_1^+W_2^-,\ W_1^-W_2^+
\label{plusminus}
\eea
proceed through the s-channel exchange of the photon, $Z_1$,  $Z_2$
or the neutral Higgses (there are alltogether six physical neutral Higgs states
in the model), and through the t-channel exchanges of the neutrinos $\nu_1$ and
$\nu_2$. Although favoured kinematically, the cross section of the latter
reaction is much smaller than that of the former because it is possible only
through the neutrino and/or weak boson mixing. The Higgs contribution is in
general negligible in both reactions \cite{higgs}.

Of a special interest is
dependence of the cross sections of the reactions (\ref{plusminus}) on the
heavy
neutrino mass $m_2$. Since there is no lepton number violation, this process
does not
directly probe the large Majorana mass term of $\nu_R$, and hence the mass
effect is not
that dramatic. Nevertheless, just above the threshold the cross section behaves
quite
differently as a function of collision energy depending on the value of $m_2$
\cite{Maalampi} (see Fig. 1). This is a quite  significant effect and may be
experimentally
detectable.

\section{Lepton number violation in $e^-e^-$ collision}

The lepton number violation associated with the triplet Higgs couplings and
Majorana
neutrinos gives rise to  many distinctive  signals of left-right symmetry. One
interesting
process is "inverse neutrinoless double beta decay"\cite{rizzo},
\cite{Maalampi},
\cite{Minkowski} \be
e^-e^-\to W^-_2W^-_2.
\label{WW}\ee
The main contributions to this reaction  comes from  the $\Delta^{--}$ exchange
in s-channel
and heavy Majorana neutrino $\nu_2$ exchange in t-channel. One should note that
any model
having only one of these particles would violate unitarity; the unitarity is
saved by a
destructive interference of the $\Delta^{--}$ exchange and $\nu_2$ exchange
amplitudes.

 The reaction (\ref{WW})
offers a probe of the symmetry breaking sector of the theory, as well as of the
nature of
the neutrinos. The most clean signal is obtained through the decay chain
$W_2\to W_1Z_1\to
3l$ +missing energy, for which there is no significant background from the
Standard Model
processes. Generally the cross section of (\ref{WW}) can be quite large, of the
order of 1
pb, as shown in Fig. 2. This would correspond to event rates of  $10^{4}$ for
an integrated
luminosity in the range of 0.1 pb$^{-1}$. The rate is about the same as that of
the
$W^+W^-$ production in the Standard Model.

Even more clear signature would have the processes \cite{Martti}
\be
e^-e^-\to\mu^-\mu^-,\ \tau^-\tau^-.
\label{muons}\ee
The two leading contributions to these reactions come from the $\Delta^{--}$
exchange in the s-channel and 		box-diagrams with virtual Majorana neutrinos
and charged
gauge bosons. Although the total lepton number is
conserved in the processes, the  lepton numbers $L_e$ and $L_{\mu}$ or
$L_{\tau}$ are
violated by two units. Also these  processes
have practically no background from the Standard Model phenomena. (The
background
opposite-sign muon pairs, produced via two-photon processes, can be separated
from the
signal by having  a magnetic field in detector.)

The cross section, which depends on the unknown masses of $\Delta^{--}$ and
$\nu_2$,
can be as high as 0.1 --1 pb, {\it i.e.}, comparable with that of the $WW$
production.  It
would be
 possible to explore quite a large range of mass values $m_N$, $M_{\Delta}$ and
$M_{W_R}$ by using this process \cite{Martti}.

\section{Tests of a susy left-right model}

The \lr\  has a naturality problem similar to that
of the Standard Model: the masses of the Higgs scalars diverge
quadratically.  As in the Standard Model, the
supersymmetry (susy) can be used to cure this hierarchy problem. To be
 theoretically satisfactory  \lr\ should be supersymmetrized (see e.g.
\cite{Ma},
\cite{Frank}).

Apart from the existence of the supersymmetric particles, the most significant
difference between the ordinary and  the supersymmetric left-right model
concerns the Higgs sector. In
supersymmetrization, the cancellation of chiral anomalies among the fermionic
partners of the triplet Higgs fields requires that  the Higgs triplet $\Delta$
is
accompanied by another triplet, $\delta$,  with opposite $U(1)_{B-L}$ quantum
number. Due to the conservation of the $B-L$ symmetry, $\delta$ does not couple
with leptons and quarks. Also  another bidoublet should be added
to avoid trivial  Kobayashi-Maskawa matrix for quarks. This comes about because
supersymmetry forbids the Yukawa coupling in which the bidoublet appears as
conjugated, so that the $u$-type quarks and $d$-type quarks should have
bidoublets of
their own (denoted by $\phi_u$ and $\phi_d$).

In \cite{kati} we have investigated a model described by the superpotential

\bea
W & = & h_u^Q \widehat Q_L^{cT} \widehat \phi_u  \widehat Q_R
+ h_d^Q \widehat Q_L^{cT} \widehat \phi_d  \widehat Q_R \nonumber \\
&&+f_u^L \widehat L_L^{cT} \widehat \phi_u  \widehat L_R
+f_d^L \widehat L_L^{cT} \widehat \phi_d  \widehat L_R
+h_R \widehat L_R^{T} i\tau_2 \widehat \DE_R  \widehat L_R \nonumber\\
&& + \mu_1 {\rm Tr} (\tau_2 \widehat \phi_u^T \tau_2 \widehat \phi_d )
+\mu_2 {\rm Tr} (\widehat \DE_R \widehat \D_R ) .\label{pot}
\eea

Here $\widehat Q_{L(R)}$ stands for the doublet of left(right)-handed quark
superfields, $\widehat L_{L(R)}$ stands for the doublet of left(right)-handed
lepton superfields, $\widehat \phi_u$ and $\widehat \phi_d$ are the two
bidoublet
Higgs superfields, and $\widehat \DE_R$ and $ \widehat \D_R$ the two
right-handed triplet
Higgs superfields. The generation indices of the quark and lepton superfields
are
not shown. It should be noticed that the mass matrix of the doubly charged
triplet
higgsinos, following from the last term of the superpotential, is particularly
simple,
because the doubly charged higgsinos do not mix with gauginos.

The next generation linear electron colliders will, besides the  $e^+e^-$
and $e^-e^-$  reactions, be able to operate also in the photon modes
 $e^-\gamma$ and $\gamma\gamma$.
The high energy photon beams can be obtained by back-scattering of intensive
laser
beam on high energy electrons. It turns out that all these collision modes may
provide
useful processes for investigation of the susy left-right model \cite{kati}
(like it does
for the susy version of the Standard Model \cite{ruckl}). Among these are the
reactions
\be
e^+e^-\to \tilde\Delta^{++}\tilde\Delta^{--}\label{reaction1}
\ee
\be e^-e^-\to \tilde\l^-\tilde\l^-\label{reaction2}
\ee
\be \gamma e^-\to \tilde e^+\tilde\Delta^{--}\label{reaction3}
\ee
\be \gamma\gamma \to \tilde\Delta^{--}\tilde\Delta^{++},\label{reaction4}
\ee
which have the common feature of the appearance of the doubly charged
higgsino(s)
$\tilde\Delta^{\pm\pm}$ in the final or  intermediate state. The fact that this
particle carries two units of electric charge and two units of lepton number
and
that it does not couple to quarks makes the processes most suitable and
distinctive test of the susy left-right model.

In large regions of the parameter space, the kinematically favoured
decay mode  of the triplet Higgsino is
$\tilde \Delta^{++}  \rightarrow  \tilde \ell^+ \ell^+ $. The simplest decay
mode of the
slepton $\tilde l$ is into an electron and the lightest neutralino (presumably
the lightest
supersymmetric particle):
\be
\tilde\l\to l\tilde\chi^0.
\ee
If kinematically allowed, the decays into final states with leptons accompanied
with
heavier neutralinos or charginos can take place in addition, but also then the
end-product
of subsequent cascade decays often is electrons plus invisible energy. The
doubly
charged triplet higgsino would thus have the following decay signature:
\be
\tilde\Delta^{--}\to l^-l^-\ +\ {\rm missing\ energy},
\label{llchannel}\ee
where $l$ can be any of $e,\ \mu$ and $\tau$ with practically equal
probabilities.
Accordingly,  the
signature of the pair production reaction (\ref{reaction1}), as well as of the
two photon
reaction (\ref{reaction4}), would be the purely leptonic final state associated
with
missing energy. The missing energy is carried by neutrinos, sneutrinos or
neutralinos.
Conservation of any separate lepton number may be broken in the visible  final
state. Such
final states are not possible in the Standard Model or in the minimal susy
model. The total
cross section for the total collision energy $\sqrt s =1 $TeV and the slepton
and higgsino
masses in the range of 100--400 GeV is about  0.5 pb. The cross section of the
reaction
(\ref{reaction4}) decreases with increasing mass of $\tilde\Delta$, but may be
as large as
10 pb \cite{kati}.

Purely leptonic final states accompanied with missing energy form also the
signals of the
reactions (\ref{reaction2}) and  (\ref{reaction3}).

\section{Conclusions}

The phenomenologically interesting characteristics of \lr\ are the new charged
weak bosons,
heavy right-handed Majorana neutrinos and doubly charged Higgses and higgsinos.
The
triplet Higgses (and higgsinos) mediate $\Delta L=2$ interactions, which give
rise to
clean and low-background signals. These features would be best benefitted in
the $e^-e^-$,
$\gamma e^-$ and $\gamma\gamma$ collision modes feasible at a linear collider
facility.

\bigskip
{\small
I am grateful to the organizers of this symposium for their kind invitation. I
thank my
collaborators in Turku, J. Vuori and A. Pietil\"a, and in Helsinki, K. Huitu
and M. Raidal.
 The work has been supported by
the Finnish Academy of Science.}

\bigskip
{\bf FIGURE CAPTION}

\noindent {\bf Figure 1.} The total cross section of the process $e^+e^-\to
W_2^-W_2^+$ as a function of the total collision energy for various values of
heavy neutrino mass $m_2$ and with $M_{W_2}=0.5$ TeV,  $M_{Z_2}=0.5$ TeV.

\noindent {\bf Figure 2.} The total cross section of the process $e^-e^-\to
W_2^-W_2^-$ as a function of the total collision energy for various values of
heavy neutrino mass $m_2$ and with $M_{W_2}=M_{Z_2}= M_{\Delta}=0.5$ TeV.

\end{document}